# Variable density preserving topology grids and the digital models for the plane

Alexander V. Evako

Volokolamskoe Sh. 1, kv. 157, 125080, Moscow, Russia
e-mail: evakoa@mail.ru.

**Abstract**

We define LCL decompositions of the plane and investigate the advantages of using such decompositions in the context of digital topology. We show that discretization schemes based on such decompositions associate, to each LCL tiling of the plane, the digital model preserving the local topological structure of the object. We prove that for any LCL tiling of the plane, the digital model is necessarily a digital 2-manifold. We show that elements of an LCL tiling can be of an arbitrary shape and size. This feature generates a variable density grid with a required resolution in any region of interest, which is extremely important in medicine. Finally, we describe a simple algorithm, which allows transforming regions of interest produced by the image acquisition process into digital spaces with topological features of the regions.

**Key words**

Topology discretization tiling tessellation cover digital model manifold graph plane

## 1 introduction

Integrating topological features into discretization and segmentation procedures in order to generate topologically correct digital models of anatomical structures is critical for many clinical and research applications [1, 3, 14]. Sometimes, particular regions of the object require a dense grid while a relatively coarse grid can be used over the rest of the object of interest. In such cases, it is suitable to use variable density grids according to external requirements and geometrical and topological features of the object.
A considerable amount of works has been devoted to building two-, three- and n-dimensional grids, e.g., [2, 11-13].
In the present paper, we use an approach, which was introduced and studied in [5-7] and was based on LCL discretization of n-dimensional objects. This type of discretization has several obvious advantages. In the discrete model (the grid), topology equivalent elements (n-tiles) are used and at the same time, the shape and the size of an individual n-tile can be arbitrary (an n-tile is not necessarily a convex set) within the framework of an LCL tiling. This allows obtaining more detailed geometrical and topological information about the regions of interest. Another feature is that the intersection graph (digital model) of the grid is a digital n-dimensional manifold preserving the topology of the object.
The material to be presented below begins with basic definitions and results related to digital objects in section 2.

We study in section 3 discretization of the plane by LCL tilings. We formulate conditions for a tiling for the plane to be the LCL cover. We show that one can choose an LCL grid with a required density in any region of interest which is extremely important in medicine. We prove that for any LCL tiling for the plane, the intersection graph is necessarily a digital 2-manifold. A trivial result of this consideration is that the quantity of non-isomorphic digital models (and LCL grids) of the plane is not restricted by a number. We provide a simple algorithm, which constructs digital models of areas of interest with any required resolution.

## 2 Preliminaries

A *digital object G* is a simple undirected graph $G=(V,W)$, where $V=\{v_1,v_2,...v_n,...\}$ is a finite or countable set of points, and $W=\{(v_pv_q),....\}\subseteq V\times V$ is a set of edges provided that $(v_pv_q)=(v_qv_p)$ and $(v_pv_p)\notin W$ [7]. Such notions as the connectedness, the adjacency, the dimensionality and the distance on a graph G are completely defined by sets V and W. Further on, if we consider a graph together with the natural topology on it, we will use the phrase *'digital space"*. We use the notations $v_p\in G$ and $(v_pv_q)\in G$ if $v_p\in V$ and $(v_pv_q)\in W$ respectively if no confusion can result.

Since in this paper we use only subgraphs induced by a set of points, we use the word *subgraph* for an induced subgraph. Points $v_p$ and $v_q$ are called *adjacent* if $(v_pv_q)\in W$. The subgraph $O(v)\subseteq G$ containing all points adjacent to v (without v) is called *the rim or the neighborhood of point v in G*, the subgraph $v\oplus O(v)$ is called *the ball of v*.

Graphs (digital spaces) can be transformed from one into another in a variety of ways. Contractible transformations of graphs [10] seem to play the same role in this approach as a homotopy in algebraic topology.

If a graph *G* is obtained from a graph *H* by a sequence of contractible transformations, then we say that *G is homotopic (or homotopy equivalent) to H.* A graph is called *contractible* if it is homotopy equivalent to a point.

Contractible transformations retain the Euler characteristic and homology groups of a graph [10]. Let us remind some necessary definitions.

A *digital 0-dimensional sphere* is a disconnected graph $S^0(a,b)$ with just two points *a* and *b*.
A connected space *M* is called *a digital n-dimensional manifold, n>0,* if the rim $O(v)$ of any point *v* is a digital *(n-1)*-dimensional sphere [4, 6].
A connected space *M* is called a *digital n-sphere, n>0,* if for any point $v\in M$, the rim $O(v)$ is a digital *(n-1)*-sphere and the space *M-v* is contractible.
For any terminology used but not defined here, see Harary [9].

## 3 LCL tilings and digital models of the plane

In this section, we use intrinsic topology of an object, without reference to an embedding space if no confusion will result.

A set D is called an *n-tile,* if it is homeomorphic to a closed unit n-dimensional cube on $R^n$. A set S is called an *n-sphere,* if it is homeomorphic to a unit n-dimensional sphere on $R^{n+1}$. We denote the interior and the boundary of an n-tile D by IntD and $\partial D$ respectively, $D=IntD\cup\partial D$. Note that the boundary $\partial D$ of an n-tile D is an (n-1)-sphere. The 0-tile D is a single point for which $\partial D=\varnothing$. If S is a circle and D is a 1-tile contained in S, then B=S-IntD is a 1-tile and $D\cap B=S^0$ is a pair of points at the ends of D and B.

**Definition 3.1.**

Let W={$D_1,D_2,…$} be a collection of n-tiles, n=1, 2.
- W is called a *locally centered collection (LC collection)* if from condition $D_{i(k)} \cap D_{i(m)} \neq \emptyset$, m≠k, m, k=1,2,...s, it follows that $D_{i(1)} \cap D_{i(2)} \cap … D_{i(s)} \neq \emptyset$.
- W is called *a locally lump collection (LL collection)* if any nonempty intersection of s distinct n-tiles is an (n-s+1)-tile: $D_{i(1)} \cap D_{i(2)} \cap … D_{i(s)} = \partial D_{i(1)} \cap \partial D_{i(2)} \cap … \partial D_{i(s)} = D^{n-s+1}$.
- W is called *a locally centered lump collection (LCL collection)* if W is a locally centered collection and a locally lump collection at the same time.

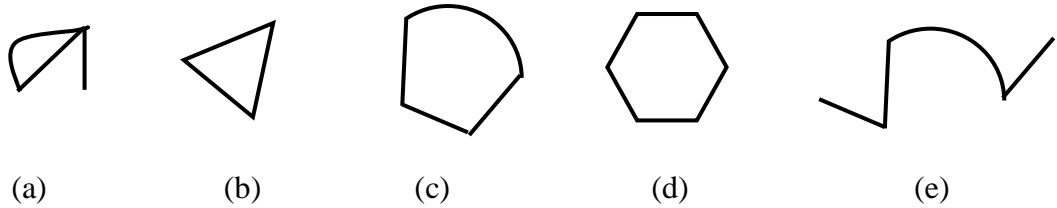

(a)  (b)  (c)  (d)  (e)

**Figure 1. Collections of 1-tile: (a) is an LC, non-LL collection. (b) is an LL, non-LC collection. (c), (d), (e) are LCL collections.**

As it follows from definition 3.1, if W={$D_1,D_2,…$} is an LCL collection of 1-tiles and $D_1 \cap D_2 \neq \emptyset$, then $D_1 \cap D_2 = \partial D_1 \cap \partial D_2 = D^0$ is a point. The intersection of three or more distinct 1-tiles is empty (fig.1).
If W={$D_1,D_2,…$} is an LCL collection of 2-tiles and $D_1 \cap D_2 \neq \emptyset$, then $D_1 \cap D_2 = \partial D_1 \cap \partial D_2 = D^1$ is a 1-tile. If $D_1 \cap D_2 \cap D_3 \neq \emptyset$, then $D_1 \cap D_2 \cap D_3 = \partial D_1 \cap \partial D_2 \cap \partial D_3 = D^0$ is a point. The intersection of four or more distinct 2-tiles is empty (fig. 2).

Evidently, an individual n-tile can be of an arbitrary shape and size within the framework of an LCL collection.
In paper [6], a locally centered collection is called continuous and it is shown that for a given object, the intersection graphs of all continuous, regular and contractible covers are

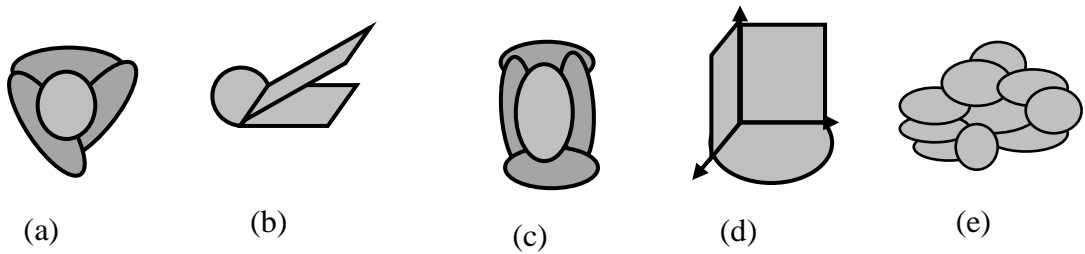

(a)  (b)  (c)  (d)  (e)

**Figure 2. Collections of 2-tiles: (a) is an LL, non-LC collection. (b) is an LC, non-LL collection. (c), (d), (e) are LCL collections.**

homotopic to each other. In paper [13], a normal set W of convex nongenerate polygons (intersection of any two of them is an edge, a vertex, or empty) is called strongly normal if for all P, $P_1,…P_n$ (n>0)∈W, if each $P_i$ intersects P and $I=P_1 \cap … \cap P_n$ is nonempty, then I intersects P (fig. 3). Several papers, e.g. [1, 11, 12] extended basic results about strong normality to collections of polyhedra in $R^n$.
There are obvious differences between SN collections of polygons and LCL collections. For example, elements of an SN collection are convex sets. On the contrary, any 2-tile in an LCL collection can be of an arbitrary form and size (fig. 2, 3) and the local topology is determined

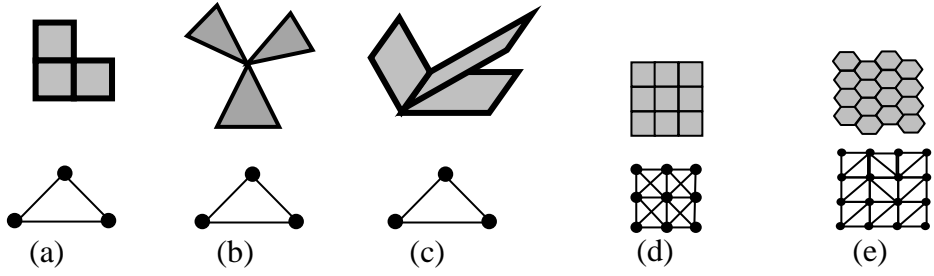

**Figure 3.** (a)-(c) are SN collections. The unions of tiles in these collections are topologically different, but the intersection graphs of the unions are identical. The collection (d) is SN but not LCL one. The collection (e) is LCL and SN one.

by the neighborhood of the tile.

The following proposition is a direct consequence of definition 3.1.

**Proposition 3.1**

(1) Let $W=\{D_0,D_1,\ldots\}$ be an LCL collection of n-tiles, n=1, 2. Then any subcollection of W is an LCL collection of n-tiles.

Regard now the set of n-tiles, n=1, 2, adjacent to a given n-tile in an LCL collection. This set specifies local topological properties of the collection.

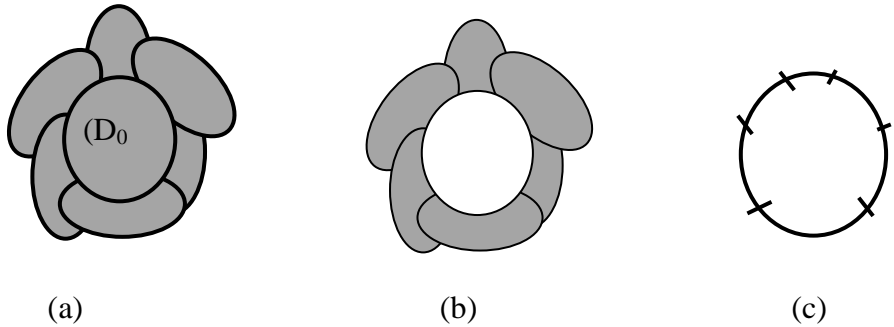

**Figure 4.** (a) An LCL collection W of 2-tiles. (b) The LCL collection U of 2-tiles adjacent to $D_0$. (c) The LCL collection V of 1-tiles $C_i=D_i \cap D_0$. Collections U and V are isomorphic.

**Proposition 3.2**

Let $W=\{D_0,D_1,\ldots\}$ be an LCL collection of n-tiles, n=1,2, and $C_i=D_0 \cap D_i \neq \emptyset$, for i=1,…s. Then the collection $V=\{C_1,C_2,\ldots C_s\}$ of (n-1)-tiles is an LCL collection and collections $U=\{D_1,D_2,\ldots D_s\}$ and $V=\{C_1,C_2,\ldots C_s\}$ are isomorphic (fig.4).

Proof.

It is obvious for n=1 (fig.1).

Let n=2. Suppose that $C_m \cap C_r \neq \emptyset$, m,r=1,…t, m≠r. Then $C_1 \cap C_2 = D_0 \cap D_1 \cap D_2 = x$ is a point. By construction, x is an endpoint of 1-tiles $C_1$ and $C_2$, i.e., $C_1 \cap C_2 = \partial C_1 \cap \partial C_2 = x$.

Assume now that $C_m \cap C_r \neq \emptyset$, m,r=1,2,3, Then $C_1 \cap C_2 \cap C_3 = D_0 \cap D_1 \cap D_2 \cap D_3 = \emptyset$. Therefore, t=2 and V is an LCL collection of 1-tiles. The isomorphism of U and V is evident. □

Fig. 4(a) shows an LCL collection $W=\{D_0,\ldots\}$ of 2-tiles, the collection U of 2-tiles adjacent to $D_0$ is depicted in fig. 4(b), the collection V of 1-tiles $C_i=D_0\cap D_i$ is shown in fig. 4(c). Collections U and V are isomorphic.

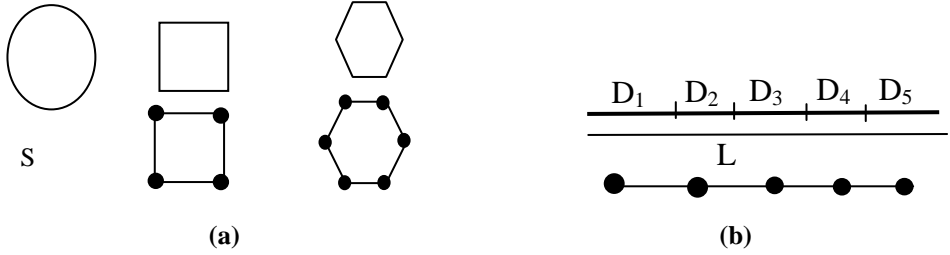

**(a)**          **(b)**

**Figure. 5. (a) LCL tilings of a circle their digital models. The intersection graphs of tilings are digital 1-spheres. (b) An LCL tiling of the line. The intersection graph of the tiling is a digital 1-manifold.**

**Definition 3.2.**

Let $W=\{D_0,D_1,...\}$ be an LCL collection of n-tiles, n=1,2. Then W is called a *tiling* of $M=D_1\cup D_2\cup....$ The intersection graph G(W) of W is called *the digital model of $M=D_0\cup D_1\cup...$ in regard to W.*

**Proposition 3.3.**
- Let an LCL collection $W=\{D_0,D_1,D_2...\}$ of 1-tiles be a tiling of a circle. Then the intersection graph G(W) of W is a digital 1-sphere (fig. 3.4).
- Let an LCL collection $W=\{D_0,D_1,D_2...\}$ of 1-tiles be a tiling of the line $R^1$. Then the intersection graph G(W) of W is a digital 1-manifold (fig. 3.4).

The proof follows from figure 5.

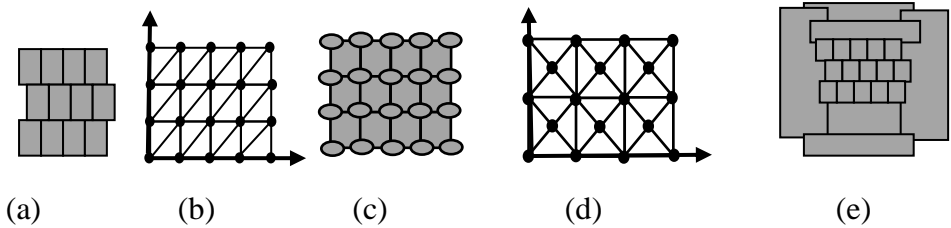

(a)      (b)      (c)      (d)      (e)

**Figure.6. LCL tilings of the plane and their digital models. (b) The digital model $Z^2$ of the tiling (a) is a digital 2-manifolds of (6,6) type. (d) The digital model $Z^2$ of the tiling (c) is a digital 2-manifold of (4,8) type. (e) The LCL tiling is a grid with variable density. In the middle of the picture the density is the highest one.**

**Proposition 3.4.**

Suppose that an LCL collection $W=\{D_0,D_1,...\}$ of 2-tiles is a tiling of the plane (fig. 6).
Then the intersection graph G(W) of W is a digital 2-manifold.

Proof.
Let $W=\{D_0,D_1,...\}$ be an LCL tiling of the plane and consider the collection $U=\{D_1,D_2...D_s\}$ of all 2-tiles, which intersect $D_0$, $D_0\cap D_i\neq\emptyset$, i=1,...s. Then the collection $V=\{C_1,C_2,...C_s\}$, where $C_i=D_0\cap D_i$, i=1,,...s, is an LCL collection of 1-tiles according to proposition 3.2. By construction, V is a tiling of the circle $S=\partial D_0$ and by proposition 3.3, the intersection graph G(V) of V is a digital 1-sphere $S^1_0$. Therefore, the intersection graph G(U) of $U=\{D_1,...D_s\}$ is a digital 1-sphere $S^1_0$. Since this is applicable to any $D_i\in W$, then $G(U_i)=S^1_i$ is a digital 1-sphere. Therefore, G(W) is a digital 2-manifold. □

As it follows from propositions 3.3. and 3.4, the digital model retains the same topology regardless of what LCL tilings are introduced.

LCL tilings of the plane and their digital models $Z^2$ are depicted in fig. 3(e) and fig. 6. Consider the LCL tilings of the plane depicted in fig. 6. In tiling (c), 2-tiles of different shape and size are used. Digital models (b) and (d) of these tilings are digital 2-manifolds $Z^2$. The tiling (e) is a grid with variable density. In the middle of the picture the density is the highest one. Note that all these tilings are not SN.

Now we are able to describe a simple algorithm, which allows transforming regions of interest produced by the image acquisition process into digital spaces with topological features of the regions.

On the first step, construct an LCL grid W with the variable density in accordance with conditions and restrictions imposed by requirements defined by the accuracy and correctness of the representation.

On the second step, build the digital model (intersection graph) of W. Using an LCL cover guaranties that G(W) is a digital 2-manifold preserving the topology of the object.

**Summary of results**

In this paper we introduce an LCL tiling for the plane and investigate its properties.
We show that the intersection graph of any LCL tiling of the plane is a digital 2-manifold preserving the local topological structure of the plane.
We present a simple algorithm for building digital counterparts of the plane with the required resolution in specific reagons of interest.

**References**

1    Y. Bai, X. Han, J. Prince: Digital Topology on Adaptive Octree Grids. Journal of Mathematical Imaging and Vision 34(2), (2009), pp. 165-184.

2    P. Brass, On strongly normal tesselations, Pattern Recognition Letters, 20(9), (1999), pp.957-960.

3    A. Dale, B. Fischl, and S. M.I. Cortical surface-based analysis i: Segmentation and surface reconstruction. NeuroImage, 9:179{194, 1999.

4    A.V. Evako, R. Kopperman, Y.V. Mukhin, Dimensional properties of graphs and digital spaces, Journal of Mathematical Imaging and Vision 6 (1996), pp. 109-119.

5    A.V. Evako, Topological properties of closed digital spaces. One method of constructing digital models of closed continuous surfaces by using covers, Computer Vision and Image Understanding 102 (2006), pp. 134-144.

6    A.V. Evako, Topological properties of the intersection graph of covers of n-dimensional surfaces, Discrete Mathematics 147 (1995), pp. 107-120.

7    A.V. Evako, The consistency principle for a digitization procedure. An algorithm for building normal digital spaces of continuous n-dimensional objects, http://www.arxiv.org/abs/math., CV/0511064, 16 Nov 2005.

8    A. V. Evako, Dimension on Discrete Spaces, International Journal of theoretical

| | |
|---|---|
| | Physics 33 (1994), pp. 1553-1568. |
| 9 | F. Harary, Graph Theory, Addison-Wesley, Reading, MA (1969). |
| 10 | A.V. Ivashchenko, Representation of smooth surfaces by graphs. Transformations of graphs which do not change the Euler characteristic of graphs, Discrete Mathematics 122 (1993), pp. 219-233. |
| 11 | T. Kong, P. Saha, A. Rosenfeld, Strongly normal sets of contractible tiles in N dimensions, Pattern Recognition 40(2) ( 2007), pp. 530-543 |
| 12 | P. K. Saha, T. Y. Kong, A. Rosenfeld, Strongly Normal Sets of Tiles in N Dimensions, Electronic Notes in Theoretical Computer Science 46 (2001), pp. 1-12. |
| 13 | P. Saha , A. Rosenfeld, Strongly normal sets of convex polygons or polyhedra, Pattern Recognition Letters, 19 (12), (1998), p.1119-1124. |
| 14 | F. Segonne, B. Fischl, Integration of Topological Constraints in Medical Image Segmentation. Biomedical Image Analysis: Methodologies and Applications, Dec 2007 |